\documentclass[review,3p,autheryear,showpacs,number,report]{elsarticle}

\usepackage{graphics}
\usepackage[colorlinks=true]{hyperref}
\usepackage{color}

\newcommand{\bw}{\begin{widetext}}
\newcommand{\ew}{\end{widetext}}
\newcommand{\be}{\begin{equation}}
\newcommand{\en}{\end{equation}}
\newcommand{\bee}{\begin{equation}}
\newcommand{\ene}{\end{equation}}
\newcommand{\bea}{\begin{eqnarray}}
\newcommand{\ena}{\end{eqnarray}}
\newcommand{\bes}{\begin{subequations}}
\newcommand{\ens}{\end{subequations}}
\newcommand{\bef}{\begin{figure}}
\newcommand{\enf}{\end{figure}}
\newcommand{\eq}[1]{Eq.~(\ref{#1})}

\def\to{\rightarrow}

\def\d{{\rm d}}

\begin{document}


\title{Deformations of the spin currents by topological screw dislocation and cosmic dispiration}

\author{Jian-hua Wang}

\author{Kai Ma\footnote{Present affiliation: KEK Theory Center and Sokendai, Tsukuba, Ibaraki 305-0801, Japan}}
\ead{makainca@gmail.com}

\address{School of Physics Science, Shaanxi University of Technology, Hanzhong 723000, Shaanxi, P. R. China}

\author{Kang Li}
\address{Department of Physics, Hangzhou Normal University, Hangzhou 310036, Zhejiang, P. R. China}

\author{Hua-wei Fan}
\address{School of Physics and Information Technology, Shaanxi Normal University, Xian 710000, Shaanxi, P. R. China}


\begin{abstract}
We study the spin currents induced by topological screw dislocation and cosmic dispiration. By using the extended Drude model, we find that the spin dependent forces are modified by the nontrivial geometry. For the topological screw dislocation, only the direction of spin current is bended by deforming the spin polarization vector. In contrast, the force induced by cosmic dispiration could affect both the direction and magnitude of the spin current. As a consequence, the spin-Hall conductivity doesn't receive corrections from screw dislocation.
\end{abstract}


\maketitle


\section{Introduction}\label{intro}
Topological defects are predicted in most of the unified theories of fundamental force. In last few decades, this subject has drawn special attention in several areas of physics ranging from condensed matter physics to cosmology \cite{Bakke:2014-1, Bakke:2014-2, Bakke:2014-3, Bakke:2013-1, Bakke:2013-2, Ma:2013:TopoDefectSHE, Ma:SHE-NCS, Chowdhury:2014-1,Chowdhury:2014-2, Chowdhury:2013-1, Chowdhury:2013-2, Chowdhury:2013-3, Chowdhury:2013-4, Galtsov:1993, Lorenci:2003, Lorenci:2002, Lorenci:2004, Montero:2011, Shen:2014, marques:landau-level, Bakke:landau-quantization, Bezerra, Bakke:gravphase, Bakke:phase-defect, Nakahara:1998, Hindmarsh:1995, kibble, Vilenkin:string, Hiscock:1985, vilenkin:1985, kleinert, dzya, Vilenkin:1983, katanaev, furtado, de-Assis:2000, furtado:landau-level, Barriola:1989, Yang, RONG1, RONG2, RONG3}. The topological defects could be formed at phase transitions in the early history of the universe, such as the cosmic string \cite{kibble, Vilenkin:string, Hiscock:1985, vilenkin:1985}, the domain wall \cite{vilenkin:1985, kleinert, dzya, katanaev, furtado, de-Assis:2000}, and the global monopole \cite{Barriola:1989}, etc. In particular, the cosmic string theory provides a bridge between the physical descriptions of microscopic and macroscopic scales, and then leads to extensive discussions on various quantum problems. The influences of topological defects on the Landau levels have been investigated in Refs.\cite{furtado:landau-level,marques:landau-level}. It was shown that the presence of a cosmic string breaks the infinite degeneracy of the Landau levels. Reference \cite{Bakke:landau-quantization} investigated the Landau quantization for a neutral particle with permanent magnetic dipole moment in the spacetime backgrounds of cosmic string and cosmic dislocation. And more recently, the relativistic and non-relativistic quantum dynamics of a neutral particle with both permanent magnetic and electric dipole moments were studied in the curved space time \cite{Bezerra}. The topological Aharonov-Bohm and Aharonov-Casher effects have also been studied in the presence of a topological defect\cite{Bakke:phase-defect, Bakke:gravphase}. Then influences on the spin-hall effects have also been studied\citep{Ma:2013:TopoDefectSHE,Ma:SHE-NCS, Chowdhury:2014-1, Chowdhury:2014-2, Chowdhury:2013-1, Chowdhury:2013-2, Chowdhury:2013-3, Chowdhury:2013-4}. Usually the nontrivial geometries could modify the spin-orbital interaction, which is the focus of this paper.

The study of spin-orbit interactions received strong attentions at ever-increasing speed since it is the theoretical foundation of the spin Hall effect (SHE) or spintronics \cite{spintronics} which studies the flow of the electron spin in the band structure of solid. The SHE was predicted in first by M. I. Dyakonov and V. I. Perel in 1971 \cite{dya1,dya2}. This effect, which occurs as a result of the spin-orbit coupling (SOC) between electrons and impurities, is called extrinsic\cite{spin-hall-hirsch}. Conversely, intrinsic mechanism also exists \cite{diffusion-zhang,band-murakami,universal-sinova,SHE-drudemodel}. It is caused by spin-orbit coupling in the band structure of the semiconductor, and then survives in the limit of zero disorder. Study of the intrinsic SHE has became an active field of research in recent years \cite{SHall1,SHall2,SHall3,ISOC,SNS,SOI}. In general, the spin current is not conserved because of the exchange of angular momentum between electron and acting electromagnetic fields through the spin-orbital interaction. Matsuo et.al. \cite{SHE-matsuo} discussed the angular momentum exchange between electron and mechanical angular momentum of the condensed matter system, and claimed that mechanical manipulation of spin currents is possible. In Ref. \cite{Ma:SHE-NCS}, SHE on noncommutative space was investigated in the first time by using the extended Drude model \cite{SHE-drudemodel}., and showed that on noncommutative space, these is a preferable direction for spin flow, and deformed accumulations of spin states on the edges of sample will occur. Based on a semiclassical approach to noncommutative quantum mechanics, SHE has also been discussed in Ref. \cite{SHE-dayi}. In this paper, we discuss the influences of a screw dislocation and a massive dispiration \cite{Galtsov:1993, Lorenci:2003} on the spin currents based on the extended Drude model.

The contents of this paper are organized as follows: In Sec. \ref{sec:dynamics}, we review the spin dynamics in the curved spacetime which has been studied in our previous paper \cite{Ma:2013:TopoDefectSHE}. In Sec. \ref{sec:dis}, we discuss the effects of the screw dislocation on the spin current and spin-Hall conductivity. It turns out that only the direction of the spin current is modified via the deformation of the polarization vector. In Sec. \ref{sec:massiveCS}, we discuss the effects of the cosmic dispiration on the spin current and spin-Hall conductivity. In this case both the direction and magnitude of the spin current receive corrections. The conclusions are given in the final Sec. \ref{conclusion}.

\section{Spin dynamics in curved spacetime}\label{sec:dynamics}
In this section, we review the dynamics of spin-1/2 particle in the electromagnetic fields in the curved space time \cite{Ma:2013:TopoDefectSHE}. In this case the Dirac equation is extended into the general convariant form\cite{Nakahara:1998},
\begin{equation}\label{C-Dirac}
    [\tilde{\gamma}^{\mu}(x)(p_{\mu}-qA_{\mu}(x)-\Gamma_{\mu}(x))+mc^2]\psi(x)=0,
\end{equation}
where $A_{\mu}$ is the electromagnetic gauge potential, $\Gamma_{\mu}(x)$ is the spinor connection, and $\tilde{\gamma}^{\mu}(x)$ are the elements of coordinate dependent Clifford algebra in the curved spacetime and satisfy the relation $\{\tilde{\gamma}^{\mu}(x),\tilde{\gamma}^{\nu}(x)\}=2g^{\mu\nu}(x)$, here $g^{\mu\nu}(x)$ is the metric of the spacetime in the presence of topological defect. The line element for a general spacetime is given by
\begin{equation}\label{line-element}
     \d s^{2}
    = g_{\mu\nu}(x) \d x^{\mu} \d x^{\nu}~.
\end{equation}
In the formalism of vierbein (or tetrad), which allows us to define the spinors in curved spacetime, the metric has the form \cite{Nakahara:1998}, 
\begin{equation}\label{eq:vierbein:def}
g_{\mu\nu}(x)=e^{a}_{~\mu}(x)e^{b}_{~\nu}(x)\eta_{ab}~,
\end{equation}
and the inverse vierbein can be defined by the relations $e^{a}_{~\mu}e^{\mu}_{~b}=\delta^{a}_{~b}$ and $e_{~a}^{\mu}e_{~\nu}^{a}=\delta^{\mu}_{~\nu}$. The spinor connection $\Gamma^{\mu}$ are connected to the vierbein with the relation
\begin{equation}\label{spinor-connection}
     \Gamma^{\mu}
    =\frac{1}{8}\omega_{\mu ab}(x)[\gamma^{a},\gamma^{b}]
    =\frac{1}{8}e_{a\nu}\nabla_{\mu}e^{\nu}_{~b}[\gamma^{a},\gamma^{b}]~.
\end{equation}
Here $\nabla_{\mu}=\partial_{\mu}+\Gamma_{\mu}$ is the covariant derivative determined by the geometry of spacetime background, and $\omega_{\mu ab}(x)$ is the one connection $\omega_{ab}(x)=\omega_{\mu ab}(x)\d x^{\mu}$. From the Generalized Dirac equation (\ref{C-Dirac}), we can get the deformed Dirac Hamiltonian as the following form,
\begin{equation}\label{re-dirac}
     H_{D}
    =\beta mc^2+c\vec{\alpha}\cdot\vec{\pi}+qA_{0}
     +\vec{\alpha}\cdot\vec{\Gamma}
     +c\vec{\alpha}\cdot\vec{\vec{\Omega}}\cdot\vec{\pi}
     +\Gamma_{0},
\end{equation}
where $\vec{\pi}=\vec p-q\vec{A}/c$ is the mechanical momentum of matter particle, and we have defined a deformation matrix for convenience,
\begin{equation}\label{re-matric}
    \Omega^{a}_{~\mu}(x)=e^{a}_{~\mu}(x)-\delta^{a}_{~\mu},~~
    \Omega^{\mu}_{~a}(x)=e^{\mu}_{~a}(x)-\delta^{\mu}_{~a}~.
\end{equation}
Moreover, the second order term $\vec{\alpha}\cdot\vec{\vec{\Omega}}\cdot\vec{\Gamma}$ in (\ref{re-dirac}) has been neglected. Comparing to the ordinary Dirac Hamiltonian there are three additional terms. $\Gamma_{0}$ behaves like an electric potential. But in our case its value is zero, and then has no influence. The term $\vec{\alpha}\cdot\vec{\Gamma}$ is directly from the spin connection of the minimal-like interaction, and behaves like a hidden momentum, which could generate a geometric phase \cite{Bakke:phase-defect}. The term $c\vec{\alpha}\cdot\vec{\vec{\Omega}}\cdot\vec{\pi}$ is induced by the geometry of the spacetime, and is determined by $g^{\mu\nu}(x)$. In this sense, it represents correction to the ordinary inner-product between $\vec\alpha$ and $\vec{\pi}$. 

The nonrelativistic approach to the dynamics of spinor in the presence of a topological defect has been study in our previous paper \cite{Ma:2013:TopoDefectSHE}, 
\begin{equation}\label{total-h}
     H_{ps}
    =H_{k}+H_{z}+H_{so}+H_{d}~.
\end{equation}
The first and final terms, $H_{k}$ and $H_{d}$ are the kinematic part with corrections of minimal coupling type and the deformed Darwin term respectively (see Ref.\cite{Ma:2013:TopoDefectSHE}). The second and third terms, $H_{z}$ and $H_{so}$ describe the Zeeman and spin-orbital interactions respectively, and the explicit expressions are
\bea
H_{z}
&=&-\frac{q\hbar}{2mc}\vec{\sigma}\cdot [\vec{B} + \vec{B}_{s} + \vec{B}_{m}]~,
\\
H_{so}
&=& \frac{q\hbar}{4m^2c^2}\vec{\sigma}\cdot\big\{ [\vec{E} +\vec{E}_{s} - \vec{E}_{m} ]\times\vec{p}\big\}~.
\ena
$\vec{B}_{s}=(\vec{\nabla}\times\vec{\Gamma})/q$ and $\vec{B}_{m}=c(\vec{\nabla}\times(\vec{\vec{\Omega}}\cdot\vec{\pi}))/q$ are the effective magnetic fields generated by the spin connection $\vec{\Gamma}$ directly and the term $\vec{\vec{\Omega}}\cdot\vec{\pi}$ which indirectly represents the geometry of the spacetime. The effective electric fields are defined as $\vec{E}_{s}=-\vec{\nabla}\Gamma_{0}/q$ and $\vec{E}_{m}=-\vec{\vec{\Omega}}\cdot\vec{\nabla}V$. The first term in $H_{so}$ describes the ordinary spin-orbital interaction and can generate a nontrivial spin current as discussed in Ref. \cite{SHE-drudemodel, Ma:2013:TopoDefectSHE}; the next two terms, which are related to the additional terms in Zeeman coupling $H_{z}$, describe the effective spin-orbital interactions and are expected to generate additional spin currents as we discussed below. In this paper we discuss only the spin-orbital interactions and the relevant Hamiltonian is,
\bee\label{soih}
H = 
 \frac{\vec{p}^2}{2m}+qV(\vec{r})
       +\frac{q\hbar}{4m^2c^2}\vec{\sigma}\cdot\bigg(\vec{E}'\times\vec{p}\bigg),
\ene
where $\vec{E}'=-(\vec{\vec{I}}-\vec{\vec{\Omega}})\cdot\vec{\nabla}V(\vec{r})$ which represents the deformation on the total electric potential $V(\vec{r})$ due to the nontrivial geometry of the spacetime. The Hamiltonian (\ref{soih}) is the general formalism of spin-orbital interaction in curved spacetime.

To discuss the dynamical consequences of this interaction, we will assume that at the leading order the ordinary Heisenberg equation is correct. Then by using the Heisenberg algebra for canonically conjugated variables $\vec{r}$ and $\vec{p}$, we have
\begin{eqnarray}
       \dot{\vec r}
    &=&\frac{\vec{p}}{m}
       +\frac{q\hbar}{4m^2c^2}\vec{\sigma}\times\vec{\nabla}V
       -\frac{q\hbar}{4m^2c^2}\vec{\sigma}\times[\vec{\vec{\Omega}}\cdot\vec{\nabla}V]\label{HAP1}
    \\
       \dot{\vec p}
    &=&-q\vec{\nabla}V(\vec{r})
       -\frac{q\hbar}{4m^2c^2}\vec{\nabla}\bigg[\bigg(\vec{\sigma}\times\vec{\nabla}V\bigg)\cdot\vec{p}\bigg]
+\frac{q\hbar}{4m^2c^2}
       \vec{\nabla}\bigg[\bigg(\vec{\sigma}\times(\vec{\vec{\Omega}}\cdot\vec{\nabla}V)\bigg)\cdot\vec{p}\bigg]\label{HAP2}
\end{eqnarray}
The third term in (\ref{HAP1}) is the cross product of the electron magnetic moment and the effective electric field in the curved space-time. From (\ref{HAP1}) we have,
\begin{equation}\label{PE1}
       \vec{p}
     = m\dot{\vec{r}}
       -\frac{q\hbar}{4mc^2}\vec{\sigma}\times\vec{\nabla}V
       +\frac{q\hbar}{4mc^2}\vec{\sigma}\times(\vec{\vec{\Omega}}\cdot\vec{\nabla}V),
\end{equation}
and then,
\begin{eqnarray}\label{PE2}
       \dot{\vec p}
    &=&m\ddot{\vec{r}}
       -\frac{q\hbar}{4mc^2}\bigg(\dot{\vec{r}}\cdot\vec{\nabla}\bigg)\bigg(\vec{\sigma}\times\vec{\nabla}\bigg)
+\frac{q\hbar}{4mc^2}
       \bigg(\dot{\vec{r}}\cdot\vec{\nabla}\bigg)\bigg(\vec{\sigma}\times(\vec{\vec{\Omega}}\cdot\vec{\nabla}V)\bigg).
\end{eqnarray}
Substituting the (\ref{PE1}) and (\ref{PE2}) into (\ref{HAP2}), one can get the dynamical equation of the canonical variable $\vec{r}$ which has the form of the Newton's second law for charge carriers,
\begin{equation}\label{dy:force}
     m\ddot{\vec{r}}
    =\vec{F}'(q, \vec{\sigma})
    =\vec{F}(q) + \vec{F}'(\vec{\sigma})
    =\vec{F}(q)+\vec{F}(\vec{\sigma})+\vec{F}_{cs}(\vec{\sigma}).
\end{equation}
Here the ordinary Lorentz force $\vec{F}(q)$ receives a correction which depends on the spin degree of freedom, $\vec{F}'(\vec{\sigma})$. It constitutes two parts: $\vec{F}(\vec{\sigma})$ which is generated by the ordinary spin-orbital interaction,
\begin{equation}\label{NFE}
     \vec{F}(\vec{\sigma})
    =-\frac{q\hbar}{4mc^2}\dot{\vec{r}}\times\bigg[\vec{\nabla}\times\bigg(\vec{\sigma}\times\vec{\nabla}V\bigg)\bigg]
     -e\vec{\nabla}V,
\end{equation}
and $\vec{F}_{cs}(\vec{\sigma})$ which results from the presence of the topological defects,
\begin{equation}\label{NFEtheta}
     \vec{F}_{cs}(\vec{\sigma})
    =\frac{q\hbar}{4mc^2}
     \dot{\vec{r}}\times\bigg[\vec{\nabla}\times\bigg(\vec{\sigma}\times(\vec{\vec{\Omega}}\cdot\vec{\nabla}V)\bigg)\bigg].
\end{equation}
Here we neglected the  terms proportional to  $1/c^4$. More interesting thing is that the total force in (\ref{dy:force}) is equivalent to a Lorentz force,
\begin{equation}\label{LFE}
     \vec{F}'(q, \vec{\sigma})
    =\frac{q}{c}\bigg(\dot{\vec{r}}\times\vec{B}'(\vec{\sigma})\bigg)-q\vec{\nabla}V(\vec{r}).
\end{equation}
which acts on a particle of charge $q$ in the electric field $\vec{E}=-\vec{\nabla}V(\vec{r})$ and magnetic field,
\begin{equation}\label{LFE}
     \vec{B}'(\vec{\sigma})
    =\vec{\nabla}\times\vec{A}'(\vec{\sigma})
    =\vec{\nabla}\times[\vec{A}(\vec{\sigma})
     +\vec{A}_{cs}(\vec{\sigma})],
\end{equation}
where
\begin{eqnarray}
       \vec{A}(\vec{\sigma})
    &=&-\frac{\hbar}{4mc}\vec{\sigma}\times\vec{\nabla}V(\vec{r}),
    \\
       \vec A_{cs}(\vec{\sigma})
    &=&\frac{\hbar}{4mc}\vec{\sigma}\times[\vec{\vec{\Omega}}\cdot\vec{\nabla}V].
\end{eqnarray}
With these knowledge, the Hamiltonian (\ref{soih}) can be rewritten as,
\begin{equation}\label{RSOIH}
     H
    =\frac{1}{2m}\bigg(\vec{p}-\frac{q}{c}\vec{A}'(\vec{\sigma})\bigg)^2
\end{equation}

By solving the equation (\ref{dy:force}), we can get the universal expression of charge and spin currents. The solution is derived by employing the extended Drude model \cite{SHE-drudemodel, Ma:2013:TopoDefectSHE} which incorporates  spin-orbit interaction into the dynamics of charge carriers. Such model allows one to obtain universal expression for spin Hall conductivity that is independent of the scattering mechanism, and then be able to show the influence of topological defect clearly.

\section{Spin currents in the presence of screw dislocation}\label{sec:dis}
In this section, we consider the dynamics of spin-1/2 particle in the electromagnetic fields in the presence of a screw dislocation\cite{Bakke:2014-1, Bakke:2014-2, Bakke:2014-3, Bakke:2013-1, Bakke:2013-2, Galtsov:1993, Lorenci:2003, Lorenci:2002, Lorenci:2004}. The line element is given by 
\begin{equation}\label{line-element}
     \d s^{2}
    =c^2\d t^{2}-\d\rho^{2}-\rho^{2}\d\varphi^{2}-(\d z + \xi\d\varphi)^{2}~,
\end{equation}
where $\xi$ is the torsion of the screw dislocation. The corresponding 
vierbeins defined in (\ref{eq:vierbein:def}) are
\begin{equation}\label{vierbein}
    e^{a}_{~\mu}(x)=
    \left(
      \begin{array}{cccc}
        1 & 0 & 0 & 0 \\
        0 & \cos\varphi & -\rho\sin\varphi & 0 \\
        0 & \sin\varphi & \rho\cos\varphi & 0 \\
        0 & 0 & \xi & 1 \\
      \end{array}
    \right)~,
\end{equation}
and,
\begin{equation}\label{vierbein}
    e^{\mu}_{~a}(x)=
    \left(
      \begin{array}{cccc}
        1 & 0 & 0 & 0 \\
        0 & \cos\varphi & \sin\varphi & 0 \\
        0 & -\frac{\sin\varphi}{\rho} & \frac{\cos\varphi}{\rho}& 0 \\
        0 & \frac{\xi}{\rho}\sin\varphi & -\frac{\xi}{\rho}\cos\varphi & 1 \\
      \end{array}
    \right)~,
\end{equation}
where the $a = (t, x, y, z), \mu = (t, \rho, \varphi, z)$. The flat spacetime can be recovered in the limit $\xi\to0$. In the rectangular coordinates, the corresponding defamation matrix defined in \eq{re-matric} is,  
\begin{equation}\label{dis:defmat}
    \Omega^{a}_{~b}= 
    \left(
      \begin{array}{cccc}
        0 & 0 & 0 & 0 \\
        0 & 0 & 0 & 0 \\
        0 & 0 & 0 & 0 \\
        0 & \frac{\xi}{\rho}\sin\varphi  & -\frac{\xi}{\rho}\cos\varphi & 0 \\
      \end{array}
    \right).
\end{equation}
In section \ref{sec:dynamics} we have reviewed the spin dynamics in a general curved spacetime. The general result is a spin dependent force, see \eq{dy:force}. In this section we would like to solve this equation for the screw dislocation, and discuss its implications. We employ the extended Drude model which is independent of the scattering mechanism to solve this equation\cite{SHE-drudemodel, Ma:2013:TopoDefectSHE}. It is worthy to note that the total electric potential $V(r)$ is the sum of the external electric potential $V_{e}(r)$ and the lattice electric potential $V_{l}(r)$. Moreover, the velocity relaxation time $\tau$ is given by experiments. We further assume that up to the first order approximation, the velocity relaxation time $\tau$ of charge carriers is independent of spin polarization. Then we can solve the equation perturbatively. The solution of \eq{dy:force} can be written in the following form \cite{SHE-drudemodel, Ma:2013:TopoDefectSHE},
\bee
\dot{\vec{r}}=\dot{\vec{r}}_{0}+\dot{\vec{r}}_{1}~,
\ene 
where $\dot{\vec{r}}_{0}$ is the solution of spin-independent part
\begin{equation}\label{FA}
     \langle\dot{\vec{r}_{0}}\rangle
    =\frac{q\tau}{m}\vec{E}~,
\end{equation}
and $\dot{\vec{r}}_{1}$ is the solution of spin- and torsion-dependent part. For a constant external electric field $\vec{E}=-\vec{\nabla}V_{e}(\vec{r})$, we can obtain the solution of $\dot{\vec{r}}_{1}$ perturbatively\cite{SHE-drudemodel, Ma:2013:TopoDefectSHE},
\bee\label{eq:current1}
\langle\dot{\vec{r}}_{1}\rangle
 = -\frac{\hbar q^2\tau^2}{4m^3c^2}
     \vec{E}\times\langle\vec{\nabla}\times [\vec{\sigma}\times[ (\vec{\vec{I}} -\vec{\vec{\Omega}})\cdot\vec{\nabla}V]]\rangle.
\ene
For cubic lattice, the electric potential could be approximated as,
\begin{equation}\label{CP}
    \langle\frac{\partial^2V_l(\vec{r})}
    {\partial r_{i}\partial r_{j}}\rangle
    =\chi\delta_{ij}~,
\end{equation}
where the constant $\chi$ have been determined in Ref. \cite{SHE-drudemodel}. In our case, \eq{eq:current1} contains the volume average of electrostatic crystal potential $\partial_i\partial_jV_l(\vec{r})$, and the derivative of the deformation matrix $\Omega_{ik}$. Motived by the observation that the real spacetime is not much deviated from the flat spacetime, we neglect the contributions that involve derivatives of  $\Omega_{ik}$. Then we obtain
\begin{equation}\label{eq:exp-r1}
\langle\dot{\vec{r}}_{1}\rangle
 =\frac{\hbar q^2\tau^2\chi}{2m^3c^2}
\bigg\{ \bigg(1-\frac{1}{2}{\rm Tr}\{\Omega\}\bigg)\vec{\sigma}
+ \frac{1}{2} (\vec{\vec{\Omega}} \cdot \vec{\sigma}  ) \bigg\} \times\vec{E}~.
\end{equation}
For the screw dislocation, the trace of the defamation matrix \eq{dis:defmat} is zero, then the correction of the velocity is
\begin{equation}\label{FAVA}
       \langle\dot{\vec{r}}_{1}\rangle
    =\frac{\hbar q^2\tau^2\chi}{2m^3c^2}[ (\vec{\vec{I}} + \frac{1}{2}\vec{\vec{\Omega}} )\cdot \vec{\sigma}  ] \times\vec{E}~.
\end{equation}
As expected the expectation value of velocity is corrected by modifying the inner product. For polarized charge carries described by the density matrix
\begin{equation}\label{SDM}
    \rho^{s}=\frac{1}{2}\rho(1+\vec{\lambda}
    \cdot\vec{\sigma}),
\end{equation}
where $\rho$ is the total concentration of charges carrying the electric current, and  $\vec{\lambda}$ is the spin polarization vector of the electron fluid. The spin current could be obtained by convolute the velocity with the density matrix and we get 
\bee
\vec{j}^{s}(\vec{\sigma}) = \sigma_{H}^{s}(\vec{\lambda}_{\xi}\times\vec{E}),
\ene
where the corresponding spin-Hall conductivity is given by,
\begin{equation}\label{SHConduct}
     \sigma_{H}^{s}
    =\frac{\hbar e^3\tau^2\rho\chi}{2m^3c^2}~,
\end{equation}
and the deformed polarization vector is
\bee\label{pv-dis}
\vec{\lambda}_{\xi} =  \bigg(\vec{\vec{I}} + \frac{1}{2}\vec{\vec{\Omega}} \bigg)\cdot \vec{\lambda}~.
\ene
Comparing to the ordinary results in Ref. \cite{SHE-drudemodel}, the spin-Hall conductivity is not affected by the screw dislocation. However, the direction of the spin current is modified. For a polarization vector parameterized as,
\bee\label{polarvector}
\vec{\lambda} = \big( \sin\bar{\theta}\cos\bar{\phi}, ~\sin\bar{\theta}\sin\bar{\phi},~ \cos\bar{\theta} \big),~
\ene
we get
\bea
\lambda_{\xi,x} &=& \sin\bar{\theta}\cos\bar{\phi} ~,
\\
\lambda_{\xi,y} &=&\sin\bar{\theta}\sin\bar{\phi}~, 
\\
\lambda_{\xi,z} &=& \sqrt{ 1 + \frac{\xi^2}{4\rho^2} \sin^2( \bar{\phi} -\varphi ) }\cos(\bar{\theta} +\hat{\theta}) \big)~,
\ena
where
\bee
\sin\hat{\theta} = \frac{ \xi \sin( \bar{\phi} - \varphi ) }{ \sqrt{ 4\rho^2 + \xi^2 \sin^2( \bar{\phi} -\varphi ) } }
\ene
For small $\xi$ we have $\hat{\theta} \approx \xi\sin( \bar{\phi} - \varphi )/(2\rho)$. The maximum deformation happens when the polarization direction has a phase difference $\pi/2$ with the momentum direction, in this case we have $\hat{\theta} = \xi/(2\rho)$. For an experimental sensitivity $\delta\theta \sim 10^{-3}$, we could investigate a screw dislocation at an order of $10^{-12} {\rm m}$ in a nanoscale system.

\section{Spin currents in the presence of cosmic dispiration}\label{sec:massiveCS}
In this section, we investigate the spin Hall effect in the presence of a cosmic dispiration. The line element is given by\cite{Bakke:2014-1, Bakke:2014-2, Bakke:2014-3, Bakke:2013-1, Bakke:2013-2, Galtsov:1993, Lorenci:2003, Lorenci:2002, Lorenci:2004}
\begin{equation}\label{line-element2}
     \d s^{2}
    =c^2\d t^{2}-\d\rho^{2}-\eta^{2}\rho^{2}\d\varphi^{2}-(\d z + \xi\d\varphi)^{2}~,
\end{equation}
where $\eta$ is the deficit angle and is defined as $\eta=1-4\lambda G/c^{2}$ with $\lambda$ being the linear mass density of the massive topological defect, $\xi$ is the torsion. The vierbeins are chossen as follow,
\begin{equation}\label{vierbein}
    e^{a}_{~\mu}(x)=
    \left(
      \begin{array}{cccc}
        1 & 0 & 0 & 0 \\
        0 & \cos\varphi & -\eta\rho\sin\varphi & 0 \\
        0 & \sin\varphi & \eta\rho\cos\varphi & 0 \\
        0 & 0 & \xi & 1 \\
      \end{array}
    \right)~,
\end{equation}
and
\begin{equation}\label{vierbein}
    e^{\mu}_{~a}(x)=
    \left(
      \begin{array}{cccc}
        1 & 0 & 0 & 0 \\
        0 & \cos\varphi & \sin\varphi & 0 \\
        0 & -\frac{\sin\varphi}{\eta\rho} & \frac{\cos\varphi}{\eta\rho}& 0 \\
        0 & \frac{\xi}{\eta\rho}\sin\varphi & -\frac{\xi}{\eta\rho}\cos\varphi & 1 \\
      \end{array}
    \right)~,
\end{equation}
where the $a = (t, x, y, z), \mu = (t, \rho, \varphi, z)$. The flat space-time can be recovered in the limit $\eta\to1$ and $\xi\to0$. In the rectangular coordinates, the corresponding defamation matrix is
\begin{equation}\label{deformmatric2}
    \Omega^{a}_{~b}= \frac{1-\eta}{\eta}
    \left(
      \begin{array}{cccc}
        0 & 0 & 0 & 0 \\
        0 & \sin^{2}\varphi & -\sin\varphi\cos\varphi & 0 \\
        0 & -\sin\varphi\cos\varphi & \cos^{2}\varphi& 0 \\
        0 & \frac{\xi}{(1-\eta)\rho}\sin\varphi & -\frac{\xi}{(1-\eta)\rho}\cos\varphi & 0 \\
      \end{array}
    \right).
\end{equation}
In section \ref{sec:dis}, based on the extended Drude model\cite{SHE-drudemodel}, we have studied the corrections on the spin currents in the presence of a screw dislocation. In this section we use the same assumptions and techniques to calculate the corrections due to a cosmic dispiration. Comparing to the case of screw dislocation, the difference is that the trace of the defamation matrix, see \eq{deformmatric2},  is non zero. We will see that this effect modifies the spin-Hall conductivity. By inserting \eq{deformmatric2} into \eq{eq:exp-r1} we get
\begin{equation}\label{eq:mcs-r1}
       \langle\dot{\vec{r}}_{1}\rangle
    =\frac{\hbar q^2\tau^2\chi (3\eta -1)}{4m^3c^2 \eta}
\bigg[  \bigg(\vec{\vec{I}} + \frac{\eta \vec{\vec{\Omega}} }{3\eta -1} \bigg)\cdot \vec{\sigma}  \bigg] \times\vec{E}~.
\end{equation}
The spin current is obtained by convoluting \eq{eq:mcs-r1} with the density matrix \eq{SDM}, and we get,
\bee\label{EJ}
\vec{j}^{s}(\vec{\sigma}, \eta) = \sigma_{H}^{s}(\eta)\bigg(\vec{\lambda}_{\xi}(\eta)\times\vec{E}\bigg),
\ene
where the spin-Hall conductivity is given by
\begin{equation}\label{SHConduct}
\sigma_{H}^{s}(\eta)
= \frac{\hbar e^3\tau^2\rho\chi (3\eta -1) }{4m^3c^2 \eta}
\approx\bigg(1+\frac{2\lambda G}{c^2}\bigg)\frac{\hbar e^3\tau^2\rho\chi}{2m^3c^2}~,
\end{equation}
and the deformed polarization vector is
\bee
\vec{\lambda}_{\xi}(\eta) = \bigg(\vec{\vec{I}} + \frac{\eta \vec{\vec{\Omega}} }{3\eta -1} \bigg) \cdot \vec{\lambda}~.
\ene
Comparing to the ordinary results in Ref. \cite{SHE-drudemodel}, and the result in last section for screw dislocation, we can see that not only the direction of spin current is deformed, the spin-Hall conductivity also receives a correction. By using the same parameterization of the polarization direction, see \eq{polarvector}, we get
\bea
\lambda_{\xi,x}(\eta) &=& \sin\bar{\theta}\cos\bar{\phi}\bigg(  1 - \frac{ (1-\eta)\sin\varphi \sin( \bar{\phi} - \varphi ) }{ (3\eta -1)\cos\bar{\phi} } \bigg)~,
\\
\lambda_{\xi,y}(\eta) &=& \sin\bar{\theta}\sin\bar{\phi}\bigg(  1 + \frac{ (1-\eta)\cos\varphi \sin( \bar{\phi} - \varphi ) }{ (3\eta -1)\sin\bar{\phi} } \bigg)~,
\\
\lambda_{\xi,z}(\eta) &=& \sqrt{1 + \frac{\xi^2 \sin^2( \bar{\phi} -\varphi ) }{(3\eta-1)^2\rho^2}  }\cos(\bar{\theta} +\hat{\theta}) ~,
\ena
with 
\bee
\sin\hat{\theta} = \frac{ \xi \sin( \bar{\phi} - \varphi ) }{ \sqrt{ (3\eta-1)^2\rho^2 + \xi^2 \sin^2( \bar{\phi} - \varphi ) } }~.
\ene
The presence of cosmic dispiration makes an important contribution to the spin Hall conductivity at an order of $\lambda G/c^2$, which has been given in Ref.\cite{Ma:2013:TopoDefectSHE}.

\section{Conclusion}\label{conclusion}
In summary, the influences of a screw dislocation and a cosmic dispiration on spin currents as well as the spin Hall conductivity have been studied. The spin dynamics is governed by the Pauli-Schrodinger Hamiltonian form which we can obtain the equation of motion of the charged particles. The Pauli-Schrodinger Hamiltonian is derived by employing the Foldy-Wouthuysen transformation which gives the general information on the non-relativistic dynamics of spin-1/2 particle in the electromagnetic fields. In the presence of topological dislocation and defect, some additional terms appear comparing to the ordinary one, see \eq{total-h}, including the corrections on the Zeeman coupling and on the spin-orbital coupling etc.. These additional terms describe the effective interactions of spin and electromagnetic fields in a flat spacetime. 

The physical consequences of these interactions are obtained by investigating the equations of motion for position operator $\vec{r}$, $m\ddot{\vec{r}}=F(q, \vec{\sigma})$, which is an quantum analogy of the second Newton's low. Furthermore, apart from the ordinary Lorentz force, there are additional Lorentz-like forces. These new Lorentz-like forces are the origin of  the corrections on spin currents and spin-Hall conductivity. Based on the extended Drude model which is independent of the scattering mechanism of the sample, the equations of motion are solved perturbatively. Comparing to the ordinary results in Ref. \cite{SHE-drudemodel}, the spin-Hall conductivity is not affected by the screw dislocation. However, the direction of the spin current is modified through the deformation of the polarization vector, see \eq{pv-dis}. For small torsion, the shift of the polar angle is  $\sim \xi\sin( \bar{\phi} - \varphi )/(2\rho)$. The maximum deformation happens when the polarization direction has a phase difference $\pi/2$ with the momentum direction, in this case we have $\hat{\theta} = \xi/(2\rho)$. For an experimental sensitivity $\delta\theta \sim 10^{-3}$, we could investigate a screw dislocation at an order of $10^{-12} {\rm m}$ in a nanoscale system. For massive topological defect, we find that not only the direction of spin current is deformed, the spin-Hall conductivity also receives a correction at an order of $\lambda G/c^2$ which has been given in Ref.\cite{Ma:2013:TopoDefectSHE}.

\noindent\textbf{Acknowledgments}: K. M. is supported by the China Scholarship Council and the Hanjiang Scholar Project of Shaanxi University of Technology.  J. H. W. is supported by the National Natural Science Foundation of China under Grant No. 11147181 and the Scientific Research Project in Shaanxi Province under Grant No. 2009K01-54 and Grant No. 12JK0960. K. L. is supported by National Natural Science Foundation of China under Grant Nos. 11175053 and 11475051.

\end{document}